\documentclass{article}
\usepackage{amsmath,graphicx,mlspconf}
\usepackage{comment}
\usepackage[table,xcdraw]{xcolor}
\usepackage{pifont}
\usepackage{url}

\copyrightnotice{U.S.\ Government work not protected by U.S.\ copyright}

\copyrightnotice{979-8-3503-2411-2/25/\$31.00 {\copyright}2025 Crown}

\copyrightnotice{979-8-3503-2411-2/25/\$31.00 {\copyright}2025 European Union}

\copyrightnotice{979-8-3503-2411-2/25/\$31.00 {\copyright}2025 IEEE}

\toappear{2025 IEEE International Workshop on Machine Learning for Signal Processing, Aug.\ 31-- Sep.\ 3, 2025, Istanbul, Turkey}

\title{Adaptable Non-parametric Approach for Speech-based Symptom Assessment: Isolating Private Medical Data in a Retrieval Datastore}
\name{Yu-Wen Chen, Julia Hirschberg}

\address{Department of Computer Science, Columbia University, United States}

\begin{document}

\maketitle

\begin{abstract}
The automatic assessment of health-related acoustic cues has the potential to improve healthcare accessibility and affordability. Although parametric models are promising, they face challenges in privacy and adaptability. To address these, we propose a NoN-Parametric framework for Speech-based symptom Assessment (NoNPSA). By isolating medical data in a retrieval datastore, NoNPSA avoids encoding private information in model parameters and enables efficient data updates. A self-supervised learning (SSL) model pre-trained on general-purpose datasets extracts features, which are used for similarity-based retrieval. Metadata-aware refinement filters the retrieved data, and associated labels are used to compute an assessment score. Experimental results show that NoNPSA achieves competitive performance compared to fine-tuning SSL-based methods, while enabling greater privacy, update efficiency, and adaptability--showcasing the potential of non-parametric approaches in healthcare.
\end{abstract}


\section{Introduction}
\label{sec:intro}

Automatic evaluation of health-related acoustic cues can improve healthcare accessibility, affordability, and scalability by enabling remote, equipment-free assessments~\cite{zimmer2022making, dang2023human}. Researchers have been increasingly exploring the potential of this field~\cite{xia2022exploring}. For example, studies have developed automated methods to diagnose {\it pertussis} from cough and whoop sounds~\cite{pramono2016cough}, detect COVID-19~\cite{shen2023piecewise}, and improve cough classification through self-supervised learning (SSL) ensembles~\cite{xue2021exploring}. In addition, acoustic health representations have been developed to enhance generalizability in respiratory health assessments and disease detection~\cite{baur2024hear, zhu2024wavrx}.

Despite these developments, the most advanced methods in acoustic assessment rely on parametric approaches, such as fine-tuning SSL models or training models from scratch~\cite{chiang2024multi,  cooper2024review}. Although effective, parametric methods pose significant challenges: they may inadvertently encode sensitive information within model parameters, raising privacy concerns in sensitive domains such as healthcare~\cite{brown2022does}; they lack flexibility in removing specific samples~\cite{guo2019certified}; and require costly retraining to adapt to new or updated data~\cite{asai2024reliable}.
To address these limitations, recent research has explored hybrid approaches that integrate parametric models with a non-parametric datastore. In~\cite{min2023silo}, the authors proposed separating training data into two distinct parts: low-risk open source data for the parametric language model (LM) and high-risk data (such as data under restrictive licenses) for the non-parametric retrieval datastore. The non-parametric datastore enables creators to easily update high-risk data from the model entirely at any time and at the level of individual examples~\cite{khandelwal2019generalization}. While non-parametric retrieval datastores have seen growing use in LMs, their application to acoustic assessment tasks remains largely unexplored. Recently,~\cite{wang2023ramp} proposed a retrieval-augmented approach to enhance synthetic voice assessment. However, this method emphasizes the integration of parametric and non-parametric components, rather than isolating assessment data for non-parametric retrieval.

\begin{figure}[htbp!]
  \centering
\centerline{\includegraphics[scale=0.85]{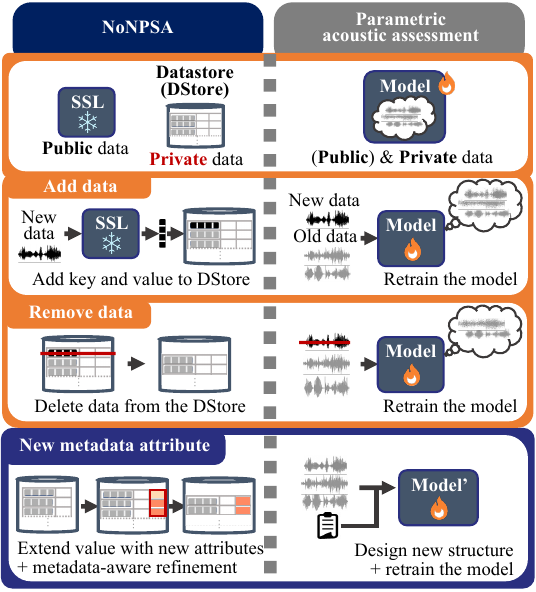}}
  \caption{NoNPSA vs. parametric acoustic assessment}
  \label{fig:compare}
\end{figure}

Inspired by~\cite{min2023silo}, we propose NoNPSA, an adaptable non-parametric speech-based symptom assessment approach that isolates private medical data within a retrieval datastore. Our study focuses on speech-based symptom assessment, determining whether a speaker exhibits respiratory symptoms based on their speech signals. First, training data is organized into a retrieval datastore, where each sample is stored as a key-value pair: the key consists of speech features extracted using a SSL model pretrained on an open-source dataset, and the value includes the label (symptomatic or asymptomatic) and associated metadata (e.g., age and sex). During inference, k-means-based segment- and utterance-level features are extracted from the input speech, and most similar samples are retrieved from the datastore to generate the assessment. This training-free approach ensures high adaptability by enabling easy updates to the data, such as adding samples or enriching existing ones with additional information, all without the need for retraining, as highlighted in the comparison with parametric models in Figure~\ref{fig:compare}. This flexibility addresses challenges posed by dynamic and incomplete medical data, making our method particularly suited for healthcare applications. Experimental results demonstrate that our method achieves competitive performance compared to parametric fine-tuned SSL-based methods, while offering the advantages of non-parametric approaches.

\section{NoNPSA}

\begin{figure*}
\centering
\centerline{\includegraphics[scale=0.8]{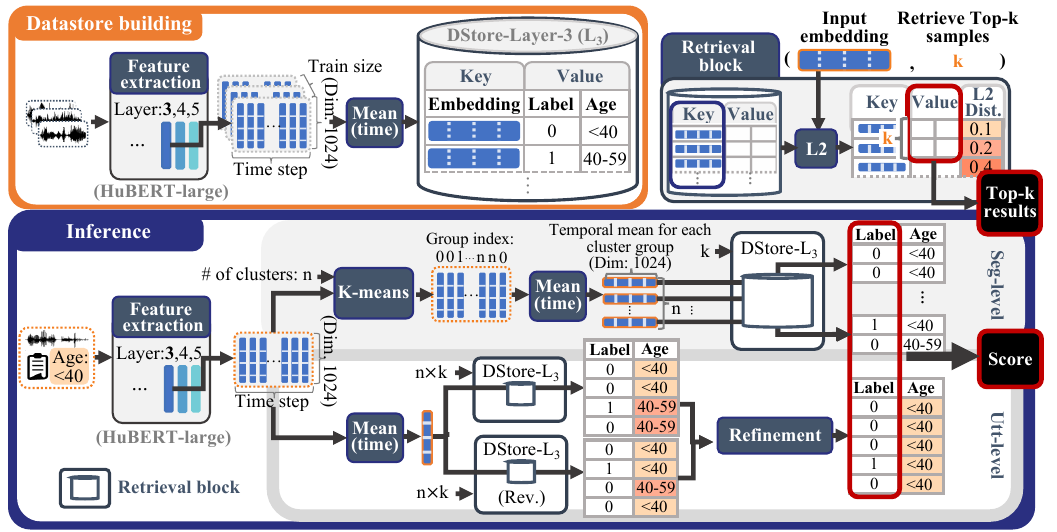}}
\caption{Overview of NoNPSA, where \emph{(Rev.)} indicates the datastore built using the reversed waveform. Note that this figure illustrates the datastore building and inference using features from layer 3 of the model. In our experiment, we repeated this process for layers \textbf{3}, \textbf{4}, and \textbf{5}, averaging the scores from these three layers to obtain the final result.}
\label{fig:proposed}%
\end{figure*}

Instead of training parametric models using symptomatic (private) data, NoNPSA leverages SSL models pretrained on a general-purpose open-source dataset to build a retrieval datastore and uses the retrieved labels to calculate assessment scores. Figure~\ref{fig:proposed} shows an overview of NoNPSA.

\subsection{Retrieval datastore}

\subsubsection{Datastore building}
Training data is organized into a retrieval datastore, with each sample stored as a key-value pair. The key represents speech features extracted using the SSL model, while the value contains the label {\it symptomatic} or {\it asymptomatic} and other metadata, such as age and sex. Each extracted feature initially has the shape ${R}^{T \times D}$, where $T$ is the number of time steps in the speech signal and $D$ is the feature dimension of the SSL model layer (e.g., $1024$ for HuBERT-large~\cite{hsu2021hubert}). A temporal mean operation is applied across the time dimension, resulting in a single vector of size ${R}^{D}$. 

We built the datastore from both the original and the reversed speech signals (i.e., with the temporal order of the waveform inverted). Since all speakers in the dataset are instructed to utter the same sentence, this study focuses solely on the acoustic characteristics of the speech signal, without considering semantic information. Although reversing the speech signal disrupts its semantic content, it provides an alternative perspective on the acoustic features and improves the performance of NoNPSA (as demonstrated in the ablation study in Section ~\ref{sec:ablation_study}). Six datastores were created using features from layer 3, 4, and 5 of the HuBERT-large model~\cite{hsu2021hubert} with both original and reversed signals for each layer. The layer selection was based on the ablation study presented in Section~\ref{sec:layer_analysis}.

\subsubsection{Retrieval process}
The input includes features extracted from the same layers as the datastore keys, along with a parameter $k$ which specifies the number of top-similar results to return. Similarity is calculated using the L2 distance between the input features and the datastore keys, and the values associated with the top-$k$ most similar keys are then retrieved.

\subsection{Inference process}
The inference process begins by extracting features using the same model layer as the datastore keys, followed by segment-level and utterance-level retrieval. In segment-level retrieval, k-means clustering~\cite{lloyd1982least} is applied to the time domain, dividing each input signal into $n$ segments. All signals from the same dataset use the same $n$ settings, as all speakers are instructed to utter the same sentence. The optimal $n$ is determined using the Silhouette score~\cite{rousseeuw1987silhouettes}, which measures clustering quality by averaging the Silhouette coefficients of all samples. Each coefficient is computed as the ratio of the difference between the mean intra-cluster distance and the mean nearest-cluster distance to the larger of the two, a score widely used for determining the optimal number of clusters. After clustering, the temporal mean for each cluster forms the segment-level features. For each of these, the top-$k$ most similar samples are retrieved from the datastore, resulting in a total of $n\times k$ retrievals ($n$ segments with $k$ samples each). 

In utterance-level retrieval, temporal averaging is applied directly to the output of the model layer. Subsequently, the top-($n\times k$) samples are retrieved from both the original and the reversed datastores, making the number of retrieved samples comparable to that in segment-level retrieval. A metadata-aware retrieval refinement is then applied to filter out samples not matching the metadata of the input. Reversed datastore retrieval and metadata-aware refinement are skipped at the segment-level to reduce complexity and prevent overfitting. Finally, labels retrieved from both segment- and utterance-level retrieval are combined to compute the assessment score. The assessment score is defined as the proportion of symptomatic samples (i.e., label 1) among all retrieved samples. An input is classified as symptomatic if the assessment score exceeds $0.5$. The process is repeated using HuBERT-large layers 3, 4, and 5. The final assessment score is derived by averaging the scores from these three layers.

\section{Experimental setup}
\subsection{Data}

We used two open-source datasets, COVID-19 Sounds~\cite{xia2021covid} (hereafter referred to as COVID-19) and Coswara~\cite{bhattacharya2023coswara}. The COVID-19 dataset contains audio recordings of participants instructed to say the sentence, “I hope my data can help manage the virus pandemic.” Each recording is labeled as either symptomatic or asymptomatic. Symptomatic recordings correspond to individuals exhibiting respiratory symptoms such as dry cough, wet cough, fever, sore throat, shortness of breath, runny nose, headache, dizziness, or chest tightness. We followed the dataset’s original split, comprising 6,648 training samples, 894 validation samples, and 1,914 test samples. For the Coswara dataset, we use the audio of participants counting from one to twenty at a normal speed. Recordings are labeled as symptomatic if the participant reports symptoms such as cough, cold, breathing difficulties, sore throat, fever, fatigue, or muscle pain. Following a 70/15/15 split for training, validation, and test sets, and after removing corrupted audio files, the dataset consists of 1,894 training samples, 402 validation samples, and 409 test samples. For both datasets, we categorize speakers into four age groups: under 39, 40–59, 60 and above, and an additional group for those with missing age information.

\subsection{Model configuration and evaluation metrics}\label{{sec:exp_setup}}

Our proposed method used hubert-large-ls960-ft\footnote{\label{note:huggingface}\url{https://huggingface.co/}} for feature extraction. For comparison, we also present results from hubert-base-ls960\footref{note:huggingface}, whisper-based models\footref{note:huggingface}, and speaker embeddings from the speechbrain spkrec-xvect-voxceleb\footref{note:huggingface}. The retrieval is implemented using the Faiss~\cite{douze2024faiss} toolkit with the IndexFlatL2 index. For COVID-19, the number of segments $n$ is set to $2$, $73$, and $73$ for layers 3, 4, and 5 of the HuBERT-large model, respectively. For the Coswara, $n$ is set to $2$ for all three layers. The k-means clustering configuration uses $n$ as the number of clusters, while all other settings follow the default configuration in scikit-learn. The segment-level retrieval parameter $k$ is set to $5$. Both $n$ and $k$ are selected based on validation set performance. Based on empirical findings presented in Section~\ref{Sec:metadata_aware}, performance on the COVID-19 is reported with age-aware refinement, whereas no refinement is applied to the Coswara. For the parametric baselines, the SSL models were fine-tuned using mean squared error loss and optimized with stochastic gradient descent with a learning rate of 0.0001 and momentum of 0.9. The best-performing model on the validation set was saved, employing an early stopping criterion with a patience of 2 epochs.

We adopt the evaluation metrics outlined in~\cite{xia2021covid}, which include:(1) ROC AUC (Receiver Operating Characteristic Area Under the Curve); (2) Sensitivity -- true positive rate or recall, defined as $TP/(TP+FN)$; and (3) Specificity, i.e., true negative rate, calculated as $TN/(TN+FP)$. The symptomatic group is considered the positive class.

\section{Results}

\subsection{Model performance}\label{sec:ablation_study}

Table~\ref{tab:model_performance} presents a performance comparison between NoNPSA and baseline parametric approaches. For the parametric methods, we fine-tuned the SSL model using a strategy proven effective for speech assessment tasks~\cite{cooper2022generalization, chen2021inqss}. Specifically, the average pooling was applied to the SSL models' output embeddings, and a dense output layer was integrated to perform symptom assessment. Results reveal that the non-parametric approach is promising for symptom assessment tasks, achieving performance competitive to parametric methods across both datasets. While our method performs better with HuBERT-large compared to HuBERT-base (Section~\ref{sec:layer_analysis}), we observe that fine-tuning HuBERT-large yields lower performance than fine-tuning HuBERT-base. A possible explanation is that available data is insufficient for effectively fine-tuning the large model. An ablation study evaluating performance using only segment-level retrieval (seg-level), or utterance-level retrieval on original (utt-level) and reversed (utt-level (Rev.)) datastore, is also presented. Results show that seg-level achieved the best ROC AUC, while utt-level (Rev.) yielded the lowest. The lower performance of utterance-level compared with segment-level may be due to a loss of finer details when averaging the embedding across all time steps. Although utt-level (Rev.) performed the lowest in terms of ROC AUC values, it achieved the highest sensitivity scores, suggesting that it provides complementary information. 


\renewcommand{\arraystretch}{1.2}
\begin{table}[th]
  \centering
 \resizebox{\linewidth}{!}{ 
\begin{tabular}{|>{\hspace{0pt}}m{0.24\linewidth}|>{\centering\hspace{0pt}}m{0.16\linewidth}|>{\centering\hspace{0pt}}m{0.13\linewidth}|>{\centering\hspace{0pt}}m{0.13\linewidth}|>{\centering\arraybackslash\hspace{0pt}}m{0.13\linewidth}|} 
\hline
\multicolumn{5}{|>{\centering\arraybackslash\hspace{0pt}}m{\linewidth}|}{{\cellcolor[rgb]{0.961,0.961,0.961}}COVID-19}    \\ 
\hline
\multicolumn{1}{|>{\centering\hspace{0pt}}m{0.24\linewidth}|}{} & Non\par{}parametric & ROC AUC & Sensitivity & Specificity \\ 
\hline
openSMILE\par{}+SVM~\cite{xia2021covid} &\ding{55}&0.63&0.56&0.62\\ 
\hline
VGGish~\cite{xia2021covid} &\ding{55}& 0.69& 0.59& 0.67 \\ 
\hline
HuBERT-base &\ding{55}& \textbf{0.708} & \textbf{0.658} & 0.758 \\ 
\hline
HuBERT-large&\ding{55}& 0.656& 0.584 & 0.727\\ 
\hline \hline
NoNPSA &\ding{51}& 0.704& 0.61  & \textbf{0.799}  \\ 
\hline
\textit{\textcolor[rgb]{0.502,0.502,0.502}{Seg-level}}          & \ding{51} & \textit{\textcolor[rgb]{0.502,0.502,0.502}{0.663}} & \textit{\textcolor[rgb]{0.502,0.502,0.502}{0.573}} & \textit{\textcolor[rgb]{0.502,0.502,0.502}{0.753}}  \\ 
\hline
\textit{\textcolor[rgb]{0.502,0.502,0.502}{Utt-level}}        &\ding{51}& \textit{\textcolor[rgb]{0.502,0.502,0.502}{0.647}} & \textit{\textcolor[rgb]{0.502,0.502,0.502}{0.552}} & \textit{\textcolor[rgb]{0.502,0.502,0.502}{0.742}}  \\ 
\hline
\textcolor[rgb]{0.502,0.502,0.502}{\textit{Utt-level (Rev.)}} &\ding{51}& \textit{\textcolor[rgb]{0.502,0.502,0.502}{0.629}} & \textit{\textcolor[rgb]{0.502,0.502,0.502}{0.599}} & \textit{\textcolor[rgb]{0.502,0.502,0.502}{0.658}}  \\ 
\hline
\multicolumn{5}{|>{\centering\arraybackslash\hspace{0pt}}m{\linewidth}|}{{\cellcolor[rgb]{0.957,0.957,0.957}}Coswara} \\ 
\hline
HuBERT-base&\ding{55}& 0.685  & 0.496 & \textbf{0.874} \\ 
\hline \hline
NoNPSA &\ding{51}& \textbf{0.723} & \textbf{0.576}& 0.870\\
\hline
\end{tabular}
}
  \caption{Performance comparison. The results for VGGish and HuBERT are from the fine-tuned models. \emph{“NoNPSA”} indicates the use of all seg-level, utt-level and utt-level (Rev.).} 
  \label{tab:model_performance}
\end{table}

\subsection{Distribution of symptom assessment scores}

\begin{figure}[htbp!]
  \centering
\centerline{\includegraphics[scale=0.9]{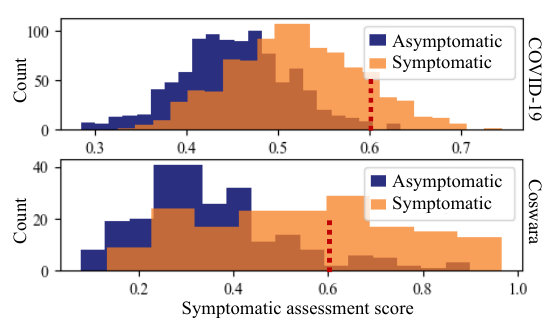}}
  \caption{Distribution of the symptom assessment score for asymptomatic and symptomatic samples.}
  \label{fig:score_distribution}
\end{figure}

Figure~\ref{fig:score_distribution} presents NoNPSA score distributions for samples labeled as asymptomatic and symptomatic. The results show that symptomatic samples generally have higher scores. Specifically, for the COVID-19, the mean and standard deviation of the scores are 0.520$\pm$0.071 for symptomatic samples and 0.452$\pm$0.060 for asymptomatic samples. For the Coswara, the corresponding values are 0.551$\pm$0.213 and 0.353$\pm$0.157, respectively. Furthermore, the distributions suggest that setting a threshold of 0.6 yields high specificity, exceeding 0.9 for both COVID-19 and Coswara, meaning that asymptomatic cases rarely have high scores.


\subsection {Analysis of metadata-aware retrieval refinement}\label{Sec:metadata_aware}

\begin{figure}[htbp!]
  \centering
\centerline{\includegraphics[scale=0.9]{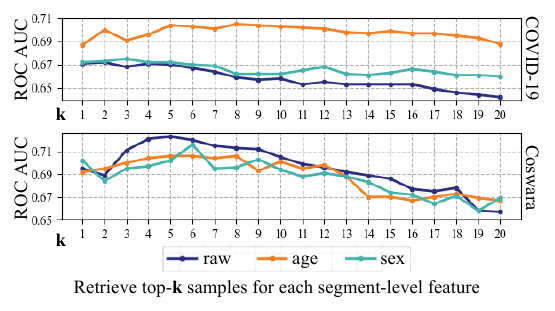}}
  \caption{Analysis of metadata-aware refinement.}
  \label{fig:refinement}
\end{figure}

We conducted an ablation study to analyze the proposed metadata-aware retrieval refinement. Figure~\ref{fig:refinement} presents the ROC AUC values for three configurations: without metadata information (denoted as raw), refining utterance-levels retrieval by age group (denoted as age), and refining by sex
group (denoted as sex). For the COVID-19, the results indicate that both age-aware and sex-aware refinement enhance retrieval performance, with age providing the most significant improvement. The smaller impact of sex may be attributed to the pretrained speech model's inherent ability to identify sex, as most retrieved samples already match the sex of the input test data. Conversely, for the Coswara, incorporating age-aware or sex-aware refinements slightly decrease performance, probably due to its smaller size -- approximately one-third that of COVID-19 -- which results in insufficient data within each metadata group. Nonetheless, the results reveal the potential of the proposed metadata-aware retrieval refinement. Unlike traditional parametric approaches that require changing model architectures and retraining with new metadata information, our method simply involves re-selecting data for retrieval, which is simple to implement and highly adaptable to new or missing metadata.

\subsection{Retrieval performance across model layers}
\label{sec:layer_analysis}

\begin{figure}[h]
  \centering \centerline{\includegraphics[scale=0.86]{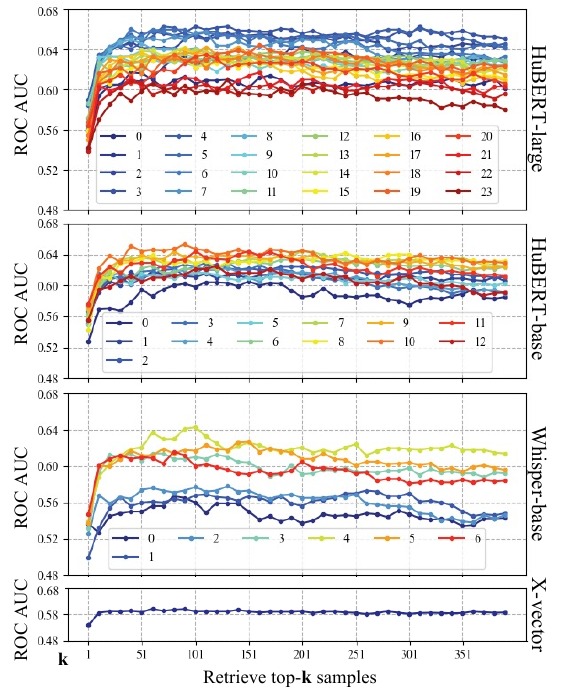}}
  \caption{Utterance-level retrieval performance on COVID-19 using features extracted from different model layers, where the legend numbers indicate the layer indices.}
  \label{fig:performance_analysis}
\end{figure}

To optimize using pre-trained SSL models for symptom assessment, we conducted an ablation study to identify the most effective model layers, including those from HuBERT-large, HuBERT-base, Whisper encoder, and x-vector~\cite{snyder2018spoken} (Figure~\ref{fig:performance_analysis}). To isolate the impact of layers and eliminate confounding factors, the experiment focused only on utterance-level retrieval using the original datastore, excluding reversed datastore results, segment-level retrieval, and refinement steps. Results reveal that earlier layers of the HuBERT-large model provided best performance. For the HuBERT-base model, the upper middle layer achieved better performance. These layers were also reported with better phoneme (acoustic) identity in previous studies~\cite{hsu2021hubert, vaidya2022self}. Whisper-base demonstrated a similar trend as HuBERT-base, where upper middle layers perform better. Overall, HuBERT models outperformed Whisper, perhaps because Whisper is primarily designed for automatic speech recognition, emphasizing semantic information. However, since speakers in our dataset were instructed to utter the same sentence, semantic information becomes less relevant for symptom assessment. Lastly, the best-performing layers from HuBERT and Whisper outperformed x-vector.

\section{Conclusion}

We propose a novel non-parametric speech-based symptom assessment (NoNPSA) framework that takes a step toward adaptable and privacy-preserving health assessments while maintaining competitive performance compared to parametric fine-tuning SSL-based methods. Our ablation study highlights specific SSL model layers that optimize performance for symptom assessment; however, the extent to which these findings generalize to other tasks remains unknown. We believe that the inherent advantages of non-parametric methods make them well-suited for healthcare applications, though their potential remains under-explored. Therefore, we plan to expand our exploration of non-parametric approaches across a wider range of health-related domains.

\bibliographystyle{IEEEbib}
\bibliography{strings,refs}

\begin{thebibliography}{10}

\bibitem{zimmer2022making}
Alexandra~J Zimmer, C{\'e}sar Ugarte-Gil, Rahul Pathri, Puneet Dewan, Devan Jaganath, Adithya Cattamanchi, Madhukar Pai, and Simon Grandjean~Lapierre,
\newblock ``Making cough count in tuberculosis care,''
\newblock {\em Communications medicine}, vol. 2, no. 1, pp. 83, 2022.

\bibitem{dang2023human}
Ting Dang, Dimitris Spathis, Abhirup Ghosh, and Cecilia Mascolo,
\newblock ``Human-centred artificial intelligence for mobile health sensing: challenges and opportunities,''
\newblock {\em Royal Society Open Science}, vol. 10, no. 11, pp. 230806, 2023.

\bibitem{xia2022exploring}
Tong Xia, Jing Han, and Cecilia Mascolo,
\newblock ``Exploring machine learning for audio-based respiratory condition screening: A concise review of databases, methods, and open issues,''
\newblock {\em Experimental Biology and Medicine}, vol. 247, no. 22, pp. 2053--2061, 2022.

\bibitem{pramono2016cough}
Renard Xaviero~Adhi Pramono, Syed~Anas Imtiaz, and Esther Rodriguez-Villegas,
\newblock ``A cough-based algorithm for automatic diagnosis of pertussis,''
\newblock {\em PloS one}, vol. 11, no. 9, pp. e0162128, 2016.

\bibitem{shen2023piecewise}
Jiakun Shen, Xueshuai Zhang, Pengyuan Zhang, Yonghong Yan, Shaoxing Zhang, Zhihua Huang, Yanfen Tang, Yu~Wang, Fujie Zhang, and Aijun Sun,
\newblock ``Piecewise position encoding in convolutional neural network for cough-based {COVID-19} detection,''
\newblock in {\em Proc. ICASSP 2023}. IEEE, pp. 1--5.

\bibitem{xue2021exploring}
Hao Xue and Flora~D Salim,
\newblock ``Exploring self-supervised representation ensembles for {COVID-19} cough classification,''
\newblock in {\em Proc. 27th ACM SIGKDD Conference on Knowledge Discovery \& Data Mining 2021}, pp. 1944--1952.

\bibitem{baur2024hear}
Sebastien Baur, Zaid Nabulsi, Wei-Hung Weng, Jake Garrison, Louis Blankemeier, Sam Fishman, Christina Chen, Sujay Kakarmath, Minyoi Maimbolwa, Nsala Sanjase, et~al.,
\newblock ``{HeAR}--health acoustic representations,''
\newblock {\em arXiv preprint arXiv:2403.02522}, 2024.

\bibitem{zhu2024wavrx}
Yi~Zhu and Tiago Falk,
\newblock ``{WavRx}: a disease-agnostic, generalizable, and privacy-preserving speech dealth diagnostic model,''
\newblock {\em IEEE Journal of Biomedical and Health Informatics}, 2024.

\bibitem{chiang2024multi}
Hsin-Tien Chiang, Szu-Wei Fu, Hsin-Min Wang, Yu~Tsao, and John~HL Hansen,
\newblock ``Multi-objective non-intrusive hearing-aid speech assessment model,''
\newblock {\em The Journal of the Acoustical Society of America}, vol. 156, no. 5, pp. 3574--3587, 2024.

\bibitem{cooper2024review}
Erica Cooper, Wen-Chin Huang, Yu~Tsao, Hsin-Min Wang, Tomoki Toda, and Junichi Yamagishi,
\newblock ``A review on subjective and objective evaluation of synthetic speech,''
\newblock {\em Acoustical Science and Technology}, pp. e24--12, 2024.

\bibitem{brown2022does}
Hannah Brown, Katherine Lee, Fatemehsadat Mireshghallah, Reza Shokri, and Florian Tram{\`e}r,
\newblock ``What does it mean for a language model to preserve privacy?,''
\newblock in {\em Proc. ACM FAccT 2022}, pp. 2280--2292.

\bibitem{guo2019certified}
Chuan Guo, Tom Goldstein, Awni Hannun, and Laurens Van Der~Maaten,
\newblock ``Certified data removal from machine learning models,''
\newblock in {\em Proc. ICML 2020}, pp. 3832--3842.

\bibitem{asai2024reliable}
Akari Asai, Zexuan Zhong, Danqi Chen, Pang~Wei Koh, Luke Zettlemoyer, Hannaneh Hajishirzi, and Wen-tau Yih,
\newblock ``Reliable, adaptable, and attributable language models with retrieval,''
\newblock {\em arXiv preprint arXiv:2403.03187}, 2024.

\bibitem{min2023silo}
Sewon Min, Suchin Gururangan, Eric Wallace, Weijia Shi, Hannaneh Hajishirzi, Noah~A Smith, and Luke Zettlemoyer,
\newblock ``{SILO} language models: Isolating legal risk in a nonparametric datastore,''
\newblock in {\em Proc. ICLR 2024}.

\bibitem{khandelwal2019generalization}
Urvashi Khandelwal, Omer Levy, Dan Jurafsky, Luke Zettlemoyer, and Mike Lewis,
\newblock ``Generalization through memorization: Nearest neighbor language models,''
\newblock in {\em Proc. ICLR 2020}.

\bibitem{wang2023ramp}
Hui Wang, Shiwan Zhao, Xiguang Zheng, and Yong Qin,
\newblock ``{RAMP}: Retrieval-augmented {MOS} prediction via confidence-based dynamic weighting,''
\newblock in {\em Proc. INTERSPEECH 2023}, pp. 1095--1099.

\bibitem{hsu2021hubert}
Wei-Ning Hsu, Benjamin Bolte, Yao-Hung~Hubert Tsai, Kushal Lakhotia, Ruslan Salakhutdinov, and Abdelrahman Mohamed,
\newblock ``{HuBERT}: Self-supervised speech representation learning by masked prediction of hidden units,''
\newblock {\em IEEE/ACM transactions on audio, speech, and language processing}, vol. 29, pp. 3451--3460, 2021.

\bibitem{lloyd1982least}
Stuart Lloyd,
\newblock ``Least squares quantization in pcm,''
\newblock {\em IEEE transactions on information theory}, vol. 28, no. 2, pp. 129--137, 1982.

\bibitem{rousseeuw1987silhouettes}
Peter~J Rousseeuw,
\newblock ``Silhouettes: a graphical aid to the interpretation and validation of cluster analysis,''
\newblock {\em Journal of computational and applied mathematics}, vol. 20, pp. 53--65, 1987.

\bibitem{xia2021covid}
Tong Xia, Dimitris Spathis, J~Ch, Andreas Grammenos, Jing Han, Apinan Hasthanasombat, Erika Bondareva, Ting Dang, Andres Floto, Pietro Cicuta, et~al.,
\newblock ``{COVID-19} sounds: a large-scale audio dataset for digital respiratory screening,''
\newblock in {\em Proc. NeurIPS 2021 datasets and benchmarks track (round 2)}, 2021.

\bibitem{bhattacharya2023coswara}
Debarpan Bhattacharya, Neeraj~Kumar Sharma, Debottam Dutta, Srikanth~Raj Chetupalli, Pravin Mote, Sriram Ganapathy, C~Chandrakiran, Sahiti Nori, KK~Suhail, Sadhana Gonuguntla, et~al.,
\newblock ``Coswara: A respiratory sounds and symptoms dataset for remote screening of {SARS-CoV-2} infection,''
\newblock {\em Scientific Data}, vol. 10, no. 1, pp. 397, 2023.

\bibitem{douze2024faiss}
Matthijs Douze, Alexandr Guzhva, Chengqi Deng, Jeff Johnson, Gergely Szilvasy, Pierre-Emmanuel Mazar{\'e}, Maria Lomeli, Lucas Hosseini, and Herv{\'e} J{\'e}gou,
\newblock ``The faiss library,''
\newblock {\em arXiv preprint arXiv:2401.08281}, 2024.

\bibitem{cooper2022generalization}
Erica Cooper, Wen-Chin Huang, Tomoki Toda, and Junichi Yamagishi,
\newblock ``Generalization ability of {MOS} prediction networks,''
\newblock in {\em Proc. ICASSP 2022}, pp. 8442--8446.

\bibitem{chen2021inqss}
Yu-Wen Chen and Yu~Tsao,
\newblock ``{InQSS}: a speech intelligibility and quality assessment model using a multi-task learning network,''
\newblock in {\em Proc. INTERSPEECH 2022}.

\bibitem{snyder2018spoken}
David Snyder, Daniel Garcia-Romero, Alan McCree, Gregory Sell, Daniel Povey, and Sanjeev Khudanpur,
\newblock ``Spoken language recognition using x-vectors.,''
\newblock in {\em Odyssey}, vol. 2018, pp. 105--111.

\bibitem{vaidya2022self}
Aditya~R Vaidya, Shailee Jain, and Alexander~G Huth,
\newblock ``Self-supervised models of audio effectively explain human cortical responses to speech,''
\newblock in {\em Proc. ICML 2022}.

\end{thebibliography}

\end{document}